\documentclass[twocolumn,showpacs,preprintnumbers,amsmath,amssymb,aps,
prb, longbibliography]{revtex4-2}
\pdfoutput=1
\usepackage{epsfig} 
\usepackage{graphicx}
\usepackage{dcolumn}
\usepackage{bm}
\usepackage{epstopdf}
\usepackage{amsmath}
\usepackage{amssymb}
\usepackage{xcolor}
\usepackage{hyperref}
\hypersetup{colorlinks = true, urlcolor = blue, linkcolor = blue} 

\begin{document}


\title{ X-ray absorption spectroscopic studies of oxygen-deficient double perovskite oxide NdBaCo$_2$O$_{5+\delta}$ ($\delta \sim 0.65$)}
\author{Himanshu Pant$^1$}
\author{S. Khalid$^2$}
\author{R. Bindu$^1$}
\altaffiliation{Corresponding author: bindu@iitmandi.ac.in}
\affiliation{$^1$School of Physical Sciences, Indian Institute of Technology Mandi, Kamand, Himachal Pradesh- 175005, India}
\affiliation{$^2$National Synchrotron Light Source-II, Brookhaven National Laboratory, Upton, New York 11973, USA}

\date{\today}

\begin{abstract}

The degree of oxygen vacancy in A-site-doped layered double perovskite oxides can lead to anion vacancy ordering, which subsequently results in distinct superstructures. We have studied oxygen deficient double perovskite oxide, NdBaCo$2$O$_{5+\delta}$ ($\delta \sim$ 0.65), utilising X-ray absorption near edge structure (XANES) spectroscopy complemented by multiple scattering and first-principles calculations based on density functional theory (DFT+U). The present results establish a direct correlation between oxygen-vacancy distribution, magnetic inhomogeneity, and charge transport in layered cobaltates. Our investigations on the Co K, Ba and Nd L$_{\textnormal{III}}$ edges, near-edge spectra based on the number and shape of the spectral features suggest that the value of $\delta$ for the compound under study lies closer to 0.75. The analysis of the Ba L$_{\textnormal{III}}$ edge spectrum confirms the +2 oxidation state for Ba ions.  The Co K-edge near-edge spectra suggest the possibility of oxygen vacancies arising at the basal sites in addition to the apical sites. This is counterintuitive to the common understanding that oxygen vacancies usually occur at the apical sites, i.e., in the lanthanide (Nd) layer in these compounds. Such distribution of oxygen vacancies naturally explains the Griffith phase and conduction mechanism of variable range hopping observed in the compound. The vacancy distribution, when at the optimum level, can also bring about easier oxygen ion mobility and a more sensitive and faster oxygen sensor. We believe that our results will help in tailoring oxygen-ion mobility and enhancing oxygen-sensing performance.

\end{abstract}

\maketitle


\section{Introduction}

In condensed matter physics, since Guatavus Rose's discovery of perovskite oxides in 1839, there has been a constant subject of investigation due to their exceptional emergent properties, which are at the heart of many scientific advancements, whether conceptual or applied. The tunability of many of the emergent properties through simple substitution makes them interesting, i.e., metal-to-insulator transitions (MIT)\cite{MIT_review_imada}, superconductivity\cite{plakida2012_Superconductivity, MIT_review_imada},magnetoresistance\cite{coey1999mixed, ramakrishnan2008strongly, taskinreviewGBCO2005}, spin density waves\cite{feng2019_SDW_CDW,charnukha2013_SDW, klauss2008_SDW, wu2001_SDW},and charge density waves \cite{MIT_review_imada} etc.


When the perovskite structure, with general formula ABO$_3$ (where A is a larger cation and B is a smaller cation), is doped at the A site or B site, then the cubic structure distorts because of the size mismatch between the dopant and original ions. At 50 \% doping at A site (B site) with A' (B'), respectively, based on their size, A and A' (B and B') ions sometimes order and stabilise in either rock salt, columnar, or layered structure\cite{king2010_cation_ordering}. These perovskites with ordered structures show doubling in the unit cells; hence, they are called \textit{double perovskites}. Double perovskites open up another degree of freedom for doping and control, showing various properties i.e. MIT\citep{maignan1999, taskinreviewGBCO2005,yamada2022physical}, 
half metallicity\citep{Half_metal_BiCu3Mn4O12,Halfmetallic_Sr2CrReO6}, ferroelectricity\cite{FE_1,FE_2, FE_3}, 
ferromagnetism, 
good ionic conductivity\cite{Ionic_conductivity_ReBaco2O6-}, magnetoresistance\cite{MR_LaCu3Mn4O12,MR_Ca_LaCu3Mn4O12_DFT,maignan1999, taskinreviewGBCO2005}  
magneto dielectric effect\cite{Magnetodielectric_Re2NiMnO6}, 
magneto caloric effect\cite{Magnetocaloric_Re2BB'O6} etc.


RBaCo$_2$O$_{5+\delta}$ (R = rare earth element) type of double perovskites comes under the class of A-site layered double perovskite oxides, where due to the smaller size of rare earth ions (R$^{3+}$) as compared to Ba$^{2+}$ ions, these oxides stabilize in a layered structure with alternating Ba-O and R-O$_\delta$ layers. This layered structure gives rise to doubling in the c-axis of the unit cell, i.e., a$_p$ $\times$ a$_p$ $\times$ 2a$_p$ type structure (112 type), where a$_p$ is equivalent to the basic perovskite unit cell\cite{maignan1999}.  Maignan\cite{maignan1999} et. al. showed that based on the lanthanide present in the system, these perovskites (prepared in air) have different degrees of oxygen vacancies in the perovskite structure; for example, as prepared DyBaCo$_2$O$_{5+\delta}$ has been reported to have $\delta\sim$ 0.3, whereas PrBaCo$_2$O$_{5+\delta}$ has been reported to have  $\delta\sim$ 0.7. Additionally, oxygen vacancy in these systems can be easily tuned during the preparation or after the preparation\cite{maignan1999, taskin2006, pralong2006_NBCO6, burley2003, solin2022}. These anion vacancies have a tendency to order in  RBaCo$_2$O$_{5+\delta}$ giving rise to different superstructures in these systems. Other than structural changes these anion vacancies can be used to tune the physical properties of this class of compounds\cite{taskinreviewGBCO2005, taskin2006} i.e. magnetic, thermoelectric, electrical and thermal transport, etc.

The crystal structure and physical properties of NdBaCo$_2$O$_{5+\delta}$ compounds are heavily dictated by their preparation conditions and oxygen non-stoichiometry ($\delta$)\cite{burley2003, zhang2023zero}. While compounds with $\delta = 0$ or $1$ without anion vacancy ordering generally stabilise in a tetragonal structure ($P4/mmm$) with a 112 superstructure\cite{burley2003, pralong2006_NBCO6}, the compound with $\delta = 0.5$ is reported to have alternating empty layers along the b-axis, creating an orthorhombic ladder structure ($Pmmm$) and a 122 superstructure. The situation becomes trickier in the $0.5 < \delta < 1$ hole-doping region; specifically, at $\delta = 0.75$, additional vacancy ordering doubles both the $a$ and $b$ directions into a 222 superstructure\cite{lobanovsky2006, khalyavin2008}. 


Additionally, in these compounds it has been commonly observed that the oxygen vacancies are present in the Nd layer, i.e., the apical oxygen positions of the $\text{CoO}_6$ octahedra \cite{maignan1999, kwon2020probing}. The size mismatch between Nd and Ba ions present at the A site generates substantial local lattice strain, which is effectively minimised by removal of apical oxygens, thereby reducing the coordination environment around the smaller $\text{Nd}^{3+}$ ions. The long-range ordering of these vacancies is well established using neutron diffraction studies for $\delta$ = 0.5 and 0.75\cite{burley2003, khalyavin2008}. However, in the case of the as-prepared sample having $\delta$ = 0.65, absorption studies will be important in understanding from where extra oxygen might be missing locally. Extended X-ray absorption fine structure (EXAFS) and X-ray absorption near-edge spectroscopy (XANES) studies will be vital to understand the vacancy positions. However, in the compound under study, there are three kinds of Co ions, all of which are distorted, i.e., highly asymmetric octahedra and square pyramids. However, the bond-length difference between apical and basal bonds is $\sim$ 0.08; that is much lower than the experimental resolution, $\Delta$R = $\pi/2(\textnormal{k}_{max} - \textnormal{k}_{min})$, for the used k range (3 \AA\ - 12 \AA) in our data. This makes EAXFS analysis extremely difficult in this system to provide information about the spatial distribution of oxygen vacancies. Hence, XANES studies with systematic multiple scattering calculations will be important to understand such a distribution. These studies will also give insights into valence state and the degree of orbital hybridisation, all of which are vital for a comprehensive understanding of these systems.

In our recent report \cite{Pant_2025}, we investigated the oxygen-deficient double perovskite NdBaCo$_2$O$_{5+\delta}$ ($\delta \sim 0.65$) using laboratory-source X-ray diffraction (XRD), revealing distinct structural signatures of a tetragonal phase ($P4/mmm$) characterised by a $2a_p \times 2a_p \times 2a_p$ superstructure typically associated with a nominal composition of $\delta = 0.75$. The system exhibits a distinct resistivity minimum around $375\text{ K}$, followed by sequential long-range ferromagnetic (FM) and ferrimagnetic (FeM) transitions at $120\text{ K}$ ($T_c$) and $60\text{ K}$, respectively. The work undertaken has shown a strong structural connection with its magnetic properties. We have observed a broad maximum in the \textit{a}-parameter around T$_c$ and a change of slope in the \textit{c}-parameter. These behaviours are directly linked with the extent of hybridisation between the Co 3\textit{d} and O 2\textit{p} states through Co-O-Co bond angles\cite{Pant_2025} which is also related to the decrement in the Nd-Nd bond distance and hence with the onset of intra-layer ferromagnetic interaction within the Co and Nd sublattices (below T$_0$, $\sim$ 260 K). The onset of interlayer ferrimagnetic interaction around $\sim$ 60 K is linked with the decrement in Nd-O bonds, leading to an increment in the hybridisation of Nd 4\textit{f} and Co 3\textit{d} states\cite{Pant_2025}. The \textit{c}-parameter is also found to increase below this temperature. We also suggest that the stabilisation of short-range magnetic ordering could be the reason for the stabilisation of variable-range hopping behaviour below the T$_0$. The signature of short-range magnetic ordering is also observed in the heat capacity measurements\cite{Pant_2025}. The above-mentioned local properties are also expected to manifest in the local structural environment and electronic states of the ions present in the compound. Towards this, we have carried out X-ray absorption near-edge spectroscopic studies at Co K, Nd and Ba L$_{\textnormal{III}}$ edges. For better understanding of the temperature-dependent properties at the local level, it is vital to identify and understand the origin of the spectral features observed in these spectra collected at 300 K. For this purpose, in this work, we discuss the X-ray near-edge absorption spectroscopic studies carried out at 300 K.

In the present study, we investigate the electronic structure of NdBaCo$_2$O$_{5+\delta}$ ($\delta\sim$ 0.65), using X-ray absorption near edge spectra at Co K, Nd L$_{\textnormal{III}}$ and Ba L$_{\textnormal{III}}$ edges, combined with first-principles DFT+U and multiple-scattering calculations. The detailed analysis of Co K and Nd L$_{\textnormal{III}}$ edges, collected at 300 K, indicates that the local structure of the compound shows the signatures close to the NBCO 575 structure. The shape of the different absorption edges aligns well with the calculated partial density of states (PDOS) using DFT+\textit{U} calculations. The Ba L$_{\textnormal{III}}$ spectrum further validates the +2-oxidation state of Ba ions; moreover, the shape and number of spectral features in the Ba L$_{\textnormal{III}}$ near-edge region are closer to those expected for a compound with $\delta$ between 0.5 and 0.75. A more detailed investigation for the Co K-edge by doing calculations for different cluster sizes indicates the possibility that, apart from the oxygen vacancies at the apical site, additional oxygen vacancies arise at the basal sites; that is counter-intuitive to the common understanding that the oxygen becomes absent from apical sites in the Nd layer in these compounds. These results provide a connection between the spatial distribution of oxygen vacancies, Griffith phase and conduction mechanism of variable range hopping observed in this compound.

\section{Experimental Details}

Polycrystalline sample of  NdBaCo$_2$O$_{5+\delta}$ was prepared using the glycine-nitrate auto combustion method ambient atmospheric conditions. High purity powders ($\geq$ 99 $\%$, Sigma Aldrich) of Nd(NO$_3$)$_3$.6H$_2$O, Ba(NO$_3$)$_2$, and Co(NO$_3$)$_3$.6H$_2$O were weighed in stoichiometric amounts and dissolved in deionized water and stirred thoroughly ($\sim$ 250 rpm) on magnetic hot plate at 150 $^{\circ}$C. Two molar weight of Glycine (C$_2$H$_5$NO$_2$) corresponding to constituent nitrates solute (Nd:Ba:Co:Glycine = 1:1:2:8) was taken and added slowly into this mixture. Following the continuous stirring, the entire precursor solution transformed into a gel as a result of solvent evaporation. The produced gel was heated at 350 $^{\circ}$C resulting in auto combustion. This black brown powder was thoroughly  mixed in a agate mortar and pestle and then kept at 300 $^{\circ}$C inside a tube furnace for 2 hours to make sure the combustion process is finished. This powder was then hand ground using agate mortar and pestle, pressed into pellets and sintered for 24 hours at 1000 $^{\circ}$C. 

The  x-ray absorption spectroscopic measurements were performed at Co K edge, Nd L$_{\textnormal{III}}$ and Ba L$_{\textnormal{III}}$ edge at beam line QAS, National Synchrotron Light Source (NSLS-II), Brookhaven National Laboratory. In the case of Co K edge, data was collected in fluorescence mode, while for Nd L$_{\textnormal{III}}$ and Ba L$_{\textnormal{III}}$, the data was collected  in transmission mode. For our analysis, we have used Athena and Artemis software, included in the DEMETER\cite{ravel2005athena} software package.

The x - ray diffraction (XRD) experiments were done using collected using Rigaku MiniFlex (600 W) with Cu-K$_\alpha$ radiation (1.54 \AA).

The temperature dependent dc susceptibility  and field dependent magnetisation measurements were done using  Quantum Design SQUID, Magnetic Properties Measurement Systems (MPMS) in temperature range 300 K to 2 K at applied magnetic fields of 100 Oe and 1000 Oe.


\section{Computational Details}

Spin-polarized calculations of density of states (DOS) were conducted with a full potential linearised augmented plane wave basis with local orbitals (LAPW + lo), as developed in the Elk code\cite{dewhurst2011_Elkcode}. We utilised the generalised gradient approximation (GGA) - PBEsol for the exchange and correlation component\cite{perdew2008_PBEsol}. The onsite Coulomb interaction term (\textit{U}) was employed to account for the significant electron-electron correlation in the Nd 4\textit{f} and Co 3\textit{d} orbitals. The values of $U_{Nd}$ were 2, 4, 6, 8, and 9 eV, while those of $U_{Co}$ were 1, 2, 3, and 4 eV. $U_{Nd}$ and $U_{Co}$ denote the onsite Coulomb interactions in the Nd 4\textit{f} orbitals and Co 3\textit{d} orbitals, respectively. The values of $U_{Nd}$ and $U_{Co}$ were 9 eV and  4 eV, respectively for the data shown in this manuscript. The experimental lattice constants and atomic positions are taken from room temperature data from Pant et. al.\cite{Pant_2025}. The atomic positions obtained experimentally were initially relaxed in the spin-polarized calculations with $U_{Nd}$ and $U_{Co}$ set to 0 eV, while maintaining the lattice parameters at the experimentally determined values; these relaxed atomic positions were thereafter utilised for further computations. The muffin-tin sphere radii of 2.3, 2.3, 1.75, and 1.65 bohr were employed for Nd, Ba, Co, and O, respectively. To conclude the self-consistent cycle, the difference in total energy was established to be smaller than 10$^{-4}$ Hartree per unit cell in the calculations. The structure was considered relaxed when the overall force was below 0.5$\times$10 $^{-3}$ Hartree per unit cell. The calculations were conducted via adaptive linear mixing with a 5$\times$5$\times$5 k-point mesh for all computations.

The absorption spectra were calculated using a multiple scattering scheme as implemented in FEFF 10\cite{kas2021advanced}. The structural information to generate the absorption spectra for NdBaCo$_2$O$_{5}$ and NdBaCo$_2$O$_{5.5}$ was obtained from Burely et al.\cite{burley2003}, where both structures have been found to stabilise in an orthorhombic crystal structure with Pmmm space group. The structural information for NdBaCo$_2$O$_{5.75}$ was taken from Pant et al. \cite{Pant_2025}, where the structure stabilises in a tetragonal structure with the P4/mmm space group.



\section{Results and discussion}

\subsection{Characterisation}

\begin{figure}

\includegraphics [width = 0.48\textwidth]{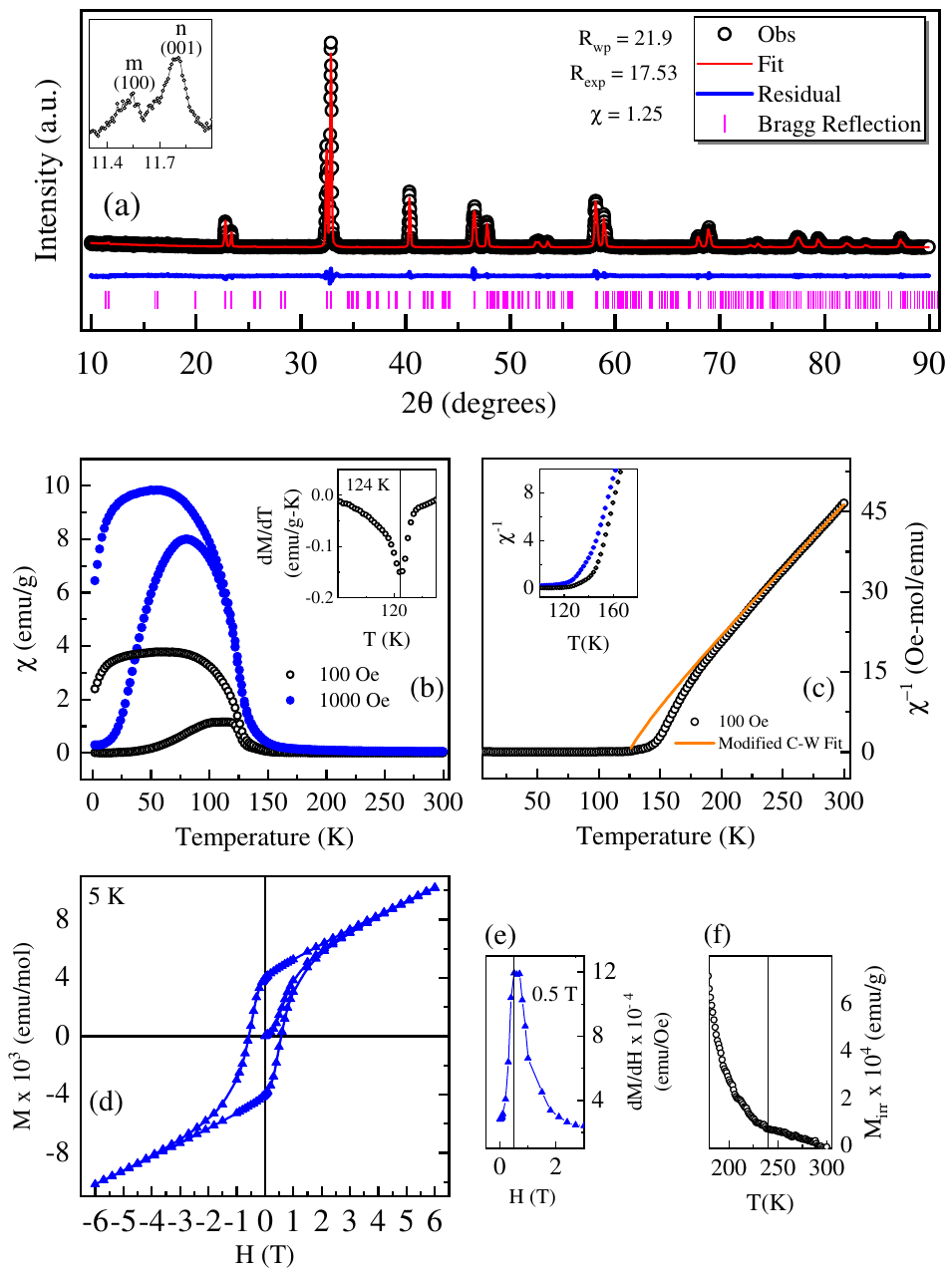}

\caption{(a) Rietveld refinement of the XRD pattern collected at room temperature; inset shows the low-angle superstructure peaks (100) and (001) after multiple slow scans. Temperature-dependent 
(b) dc susceptibility at an applied field of 100 Oe and 1000 Oe during ZFC and FC cycles; the inset shows the derivative of magnetisation with respect to field; 
(c) inverse molar susceptibility at a 100 Oe applied field during FC cycle and extrapolated modified C-W fit till T$_c$; the inset shows inverse molar susceptibility at an applied field of 100 Oe and 1000 Oe during FC cycle in 100 K. 
(d) magnetisation vs applied magnetic field at 5 K; 
(e) magnetisation with respect to field;
(f) sharp upturn in M$_{irr}$ around 240 K.}

\label{Characterisation}

\end{figure}

The Rietveld refinement of the room-temperature x-ray diffraction pattern for polycrystalline double perovskite NdBaCo$_2$O$_{5+\delta}$ is shown in Fig. \ref{Characterisation}(a). The compound is found to stabilise in a tetragonal structure with the P4/mmm space group. The inset shows superstructure peaks (100) and (001).  This suggests that the compound stabilises in 2a$_p$ $\times$ 2a$_p$ $\times$ 2a$_p$\cite{Pant_2025} (222) superstructure due to additional oxygen vacancy ordering, where a$_p$ corresponds to the basic perovskite lattice parameter. The obtained lattice parameters are a = 7.8008(3) \AA, and c = 7.6106(3) \AA, comparable with our recent report\cite{Pant_2025}, where the sample was synthesised by the solid-state reaction method. 


The temperature-dependent dc susceptibility, recorded during the zero field cooled (ZFC) and field cooled (FC) cycles at applied fields of 100 Oe and 1000 Oe, is illustrated in Fig. \ref{Characterisation}(b). The susceptibility behaviour is in line with the literature \cite{solin2022, Pant_2025}. The critical temperature (T$_c$) is derived from the first derivative of susceptibility with respect to temperature, yielding a value of approximately 124 K for an applied field of 100 Oe, shown in inset of Fig.  \ref{Characterisation}(b). Based on the obtained T$_c$ value, the $\delta$ value is estimated to be approximately 0.65 according to the literature\cite{solin2022, Pant_2025}.


In line with our previous work, we observe the signature of Griffith's phase below $\sim$ 260 K, the paramagnetic (PM) to ferromagnetic (FM) transition at 124 K and FM to ferrimagnetic (FeM) transition $\sim$ 60 K. Fig. \ref{Characterisation}(c), shows the inverse susceptibility, where the susceptibility data was fitted with the modified Curie-Weiss equation (270 K - 300 K) given as,

\begin{equation}
  \chi(T)= \dfrac{C{_{{Nd}}}}{T-\theta{_{{Nd}}}} + \dfrac{C{_{{Co}}}}{T-\theta{_{{Co}}}}
\label{CW}
\end{equation}

C${_{{Nd}}}$ and C${_{{Co}}}$ denote the Curie constants, while $\theta{_{{Nd}}}$ and $\theta{_{{Co}}}$ represent the Curie temperatures for the Nd and Co ions, respectively. Here, we assume that the C${_{{Co}}}$ associated with octahedral and square pyramidal environments are identical. During the fitting process, C${_{{Nd}}}$ and $\theta{_{{Co}}}$ were held constant at their known values of 1.62 and 124 K, respectively, resulting in C${_{{Co}}}$ = 2.60(1) and $\theta{_{{Nd}}}$ = 60.04(4) K as elaborately described in our recent report\cite{Pant_2025}. The obtained value of $\theta{_{{Nd}}}$ suggests ferromagnetic (FM) ordering among Nd ions. It is noted that as per literature\cite{khalyavin2008}, these ions are ferrimagnetic (FeM) aligned with the Co ions. The temperature dependence of the irreversible magnetisation, defined as $M_{\text{irr}} = M_{\text{FC}} - M_{\text{ZFC}}$, was plotted in the range 200-300 K, as shown in Fig. \ref{Characterisation}(f). The abrupt increase in the $M_{\text{irr}}$ value is in coincidence with the onset of the Griffiths phase characterised by the deviation of inverse susceptibility from the modified Curie-Weiss equation. 



The field-dependent magnetisation behaviour for the sample is shown in Fig.\ref{Characterisation}(d) at 5 K. The hysteresis in the sample points out the resultant FM response of the sample at low temperature. As both FM and FeM behaviours are present in the sample, a metamagnetic transition is expected in the sample. The signature of such transition is given in Fig.\ref{Characterisation}(e). The peak observed at 0.5 T distinctly indicates the change from FeM behaviour to FM behaviour, hence demonstrating the metamagnetic transition. Overall, the sample prepared via the auto-combustion method shows structural and magnetic behaviour similar to the sample prepared via the solid-state reaction route\cite{Pant_2025}. The physical properties are closely linked with the local electronic states and structural parameters around the constituent ions. We have carried out X-ray absorption spectroscopy at Nd L$_{\textnormal{III}}$, Ba L$_{\textnormal{III}}$ and Co K edges collected at 300 K. Understanding the origin of these features will be helpful in developing the link between the temperature-dependent local structural behaviour and the physical properties.







\subsection{X-ray absorption spectroscopic studies}



\subsubsection{Nd L$_{\textnormal{III}}$ edge}

\begin{figure}

\includegraphics [width = 0.5\textwidth]{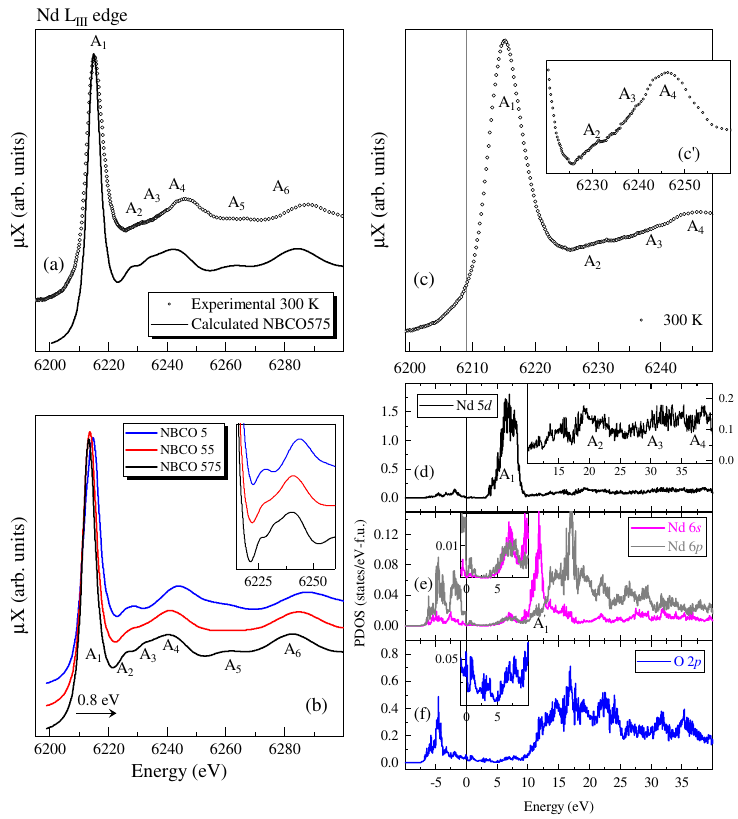}

\caption{Comparison of (a) normalised experimental Nd L$_{\textnormal{III}}$ edge at 300 K for sample under study along with calculated XANES for NBCO 575; (b) calculated XANES for NBCO 5, NBCO 55 and NBCO 575; the inset shows the magnified view of features in the energy range 6215 - 6260 eV; (c) experimental spectrum at 300 K; (c') shows a magnified view of the experimental spectrum at 300 K in 6220 - 6260 eV range. Partial density of states (PDOS) calculated using the DFT + \textit{U} approach for NBCO 575 corresponding to (d) Nd 5\textit{d}  states; the inset shows a magnified view for Nd 5\textit{d} states in 10 - 40 eV range; (e) Nd 6\textit{s} and 6\textit{p }states; the inset shows the magnified view in the (-1) - 10 eV range; and (f) O 2\textit{p} states; the inset shows the magnified view in the (-1) - 10 eV range). The calculated spectra in (a) and (b) for NBCO 5, NBCO 55, and NBCO 575 have been shifted 0.8 eV towards higher energy to approximately match the white line in the experimental spectra. In (d-f), 0 marks the Fermi level.}
\label{XANES_Nd}

\end{figure}

Fig. \ref{XANES_Nd}(a) shows the experimental absorption spectrum corresponding to the Nd L$_{\textnormal{III}}$ edge. All six observed features in the spectrum have been marked from A$_1$ to A$_6$. To explore the origin of these spectral features, the absorption spectra were calculated utilising the multiple-scattering scheme as implemented in FEFF 10 code\cite{FEFF10_kas2021}. The stabilisation of the studied compound in various crystal structures is dependent upon the oxygen stoichiometry\cite{maignan1999, taskin2006, pralong2006_NBCO6, burley2003, solin2022, Pant_2025}. To understand the local environment of the associated crystal structure, the spectra were calculated for  $\delta$ values of 0 (NBCO 5), 0.5 (NBCO 55), and 0.75 (NBCO 575). All the calculated spectra were shifted to match the white line of the experimental spectrum, with the purpose to draw a correspondence with the experimental spectrum.


 In comparison among the calculated spectra corresponding to NBCO 5, 55, and 575, we can observe that in the case of NBCO 5 and 55, there are only two features next to the white line (A$_1$), while in the case of the NBCO 575 structure, there are three features, in line with the experimental spectra shown in the inset to Fig. \ref{XANES_Nd}(b). These three features can be clearly seen in the experimental spectra in Fig. \ref{XANES_Nd}(c'). If we have a closer look at feature A$_5$, in the case of NBCO 55, this feature is not that prominent, and in the case of NBCO 5, this feature is closer to the peak equivalent to the A$_4$ peak in experimental spectra, but in the case of NBCO 575, this feature can be observed in the middle of feature A$_4$ and A$_6$, which is more in line with the experimental spectrum, Figs. \ref{XANES_Nd} (a) and (b). When we compare the calculated XANES of all three compounds as shown in Fig. \ref{XANES_Nd}(b), our results show that with regard to the number and shape of the features, the calculated spectrum of NBCO575 are very close to the experimental spectrum. This is very clearly depicted in the inset of Fig. \ref{XANES_Nd}(b).


 It is interesting to note that, in the temperature-dependent absorption studies at Nd L$_{\textnormal{III}}$ edge by Hidaka et. al.\cite{hidaka2005correlation} on NdBaCo$_2$O$_{5.5}$, near the white line (A$_1$) there are only two features; that is in line with our calculated spectra, Fig. \ref{XANES_Nd}(b).

 To shed some light on the origin of the different features close to the edge, we now compare  DFT + \textit{U} based spin-polarised calculations performed on NdBaCo$_2$O$_{5.75}$ having a tetragonal (P4/mmm) structure\cite{Pant_2025}. 

In the case of the Nd L$_{\textnormal{III}}$ edge, under the dipole approximation, during the absorption process, the allowed transition involves transitions from Nd 2\textit{p}$_{3/2}$ states to Nd 5\textit{d} and 6\textit{s}-derived states. In Fig. \ref{XANES_Nd}(c), we show the experimental Nd L$_{\textnormal{III}}$ near edge absorption spectrum, and in Fig. \ref{XANES_Nd}(d) we show the unoccupied partial density of states (PDOS) for Nd 5\textit{d} states. It is important to note that Co 3\textit{d} states also contribute above the Fermi level. Due to the localised nature of Co 3\textit{d} and it being at another site, its contribution to the absorption process is negligible. We chose the $\epsilon_F$ in the experimental absorption spectrum (6208.96 eV) by assuming the peak of the 5\textit{d} PDOS to be equivalent to the white line in the absorption spectrum.  The features A$_2$, A$_3$, and A$_4$ in the experimental absorption spectrum align well with unoccupied PDOS for Nd 5\textit{d} states, Figs. \ref{XANES_Nd} (c) and (d). Although Nd 2\textit{p}$_{3/2}$ to Nd 6\textit{s} transition probability is low, the asymmetry of feature A$_1$ indicates that 6\textit{s} states are also taking part in the absorption process, labelled as feature A$_1^{'}$, Fig. \ref{XANES_Nd} (e). Figs. \ref{XANES_Nd} (e) and (f) show the Nd 6\textit{p} and O 2\textit{p} PDOS, respectively. Around the A$_1$ feature, both the spectra show small contributions and are hybridised with Nd 5\textit{d} states above the $\epsilon_F$, as shown in the inset of Figs. \ref{XANES_Nd} (e) and (f). Nd 6\textit{p} and O 2\textit{p} PDOS show a strong and hybridised feature around A$_2$. Around A$_3$ and A$_4$ features in the experimental spectrum as well, the contribution from Nd 6\textit{p} and O 2\textit{p} states is considerable to the Nd 5\textit{d} states. This suggests that features A$_2$, A$_3$ and A$_4$ have strong contributions from \textit{p}-symmetry states along with Nd 5\textit{d}  and 6\textit{s} states showing strong \textit{p}-\textit{d} hybridisation in the compound under study. Overall, our DFT+$U$ calculations successfully explain and capture the shape of the Nd $L_{\text{III}}$ edge spectral features.



\begin{figure}

\includegraphics [width = 0.48\textwidth]{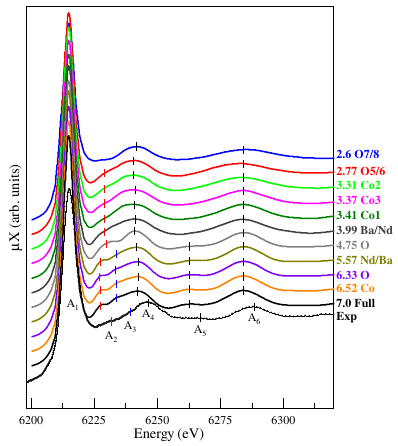}

\caption{Comparison of calculated Nd L$_\textnormal{III}$ absorption spectra for different cluster sizes with respect to the absorbing atom along with experimental spectra. The cluster size in \AA\ is given against each of the calculated spectra. The calculated spectra are shifted 0.8 eV towards higher energy, and vertical marks correspond to different features (A$_1$ - A$_6$) in the absorption spectra.}
\label{Var_Nd}

\end{figure}


To better understand the origin of distinct features (A$_1$ - A$_6$) in Nd L$_{\textnormal{III}}$ absorption spectra, multiple scattering calculations were conducted utilising various cluster sizes surrounding the Nd central atom. Figure \ref{Var_Nd} shows the calculated absorption spectra with varying cluster size in the tetragonal structure (P4/mmm) with a 222 superstructure. The features A$_1$, A$_4$ and A$_6$ start to develop with the inclusion of O and then further grow in intensity with the inclusion of other atoms. This suggests that Nd 5\textit{d} states get modified due to their hybridisation with O 2\textit{p}-symmetry states, giving rise to these features. Feature A$_2$ starts developing after the inclusion of O5 and O6 (apical oxygen present on Co2 and Co3 octahedra in the Nd layer) and grows after including other Co and O atoms. This indicates that this feature is associated with in-plane oxygens. The features A$_3$ and A$_5$ are special in the sense that they are growing in intensity only after the inclusion of the second nearest Nd and Ba atoms present around 5.57 $\AA$.


\subsubsection{Ba L$_{\textnormal{III}}$ XANES spectra}

\begin{figure}

\includegraphics [width = 0.5\textwidth]{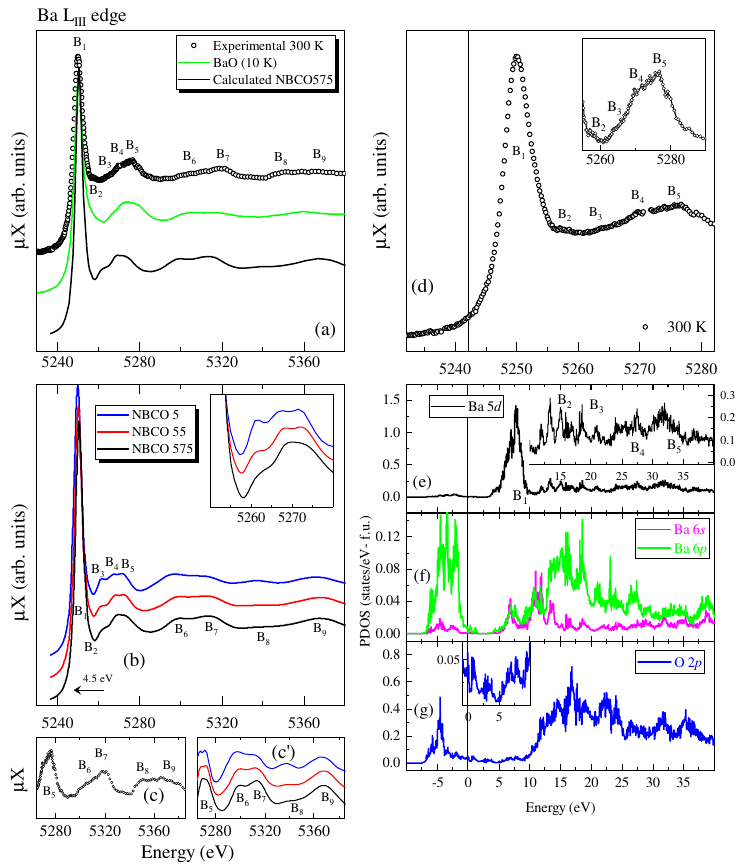}

\caption{Comparison of (a) normalised experimental Ba L$_{\textnormal{III}}$ edge at 300 K for the sample under study along with calculated XANES for NBCO 575; (b) calculated XANES for NBCO 5, NBCO 55 and NBCO 575; the inset shows
the magnified view of features in the energy range 5250 - 5280 eV; (c) magnified view of normalised experimental spectrum at 300 K (5265 - 5385 eV); (c') magnified view of calculated XANES for NBCO 5, NBCO 55 and NBCO 575 (5265 - 5385 eV); (d) experimental spectrum at 300 K; the inset shows a magnified view of the features in the energy range 5255 - 5290 eV. PDOS calculated using the DFT + \textit{U} approach for NBCO 575 corresponding to (e) Ba 5\textit{d} states; the inset shows a magnified view in 10 - 40 eV range; (f) Ba 6\textit{s} and 6\textit{p }states; and (g) O 2\textit{p} states; the inset shows the magnified view in the (-1) - 10 eV range. The calculated spectra in (a), (b) and (c') for NBCO 5, NBCO 55, and NBCO 575 have been shifted 4.5 eV towards lower energy to match the white line in the experimental spectra. In (d-f), 0 marks the Fermi level. }
\label{XANES_Ba}

\end{figure}

Fig. \ref{XANES_Ba}(a) shows the experimental absorption spectra corresponding to the Ba L$_{\textnormal{III}}$ edge at 300 K for NdBaCo$_2$O$_{5+\delta}$ and at 10 K for the BaO sample. If we compare the white line position in the case of the sample under study and BaO, it is almost the same, indicating that the oxidation state for Ba is +2 in the studied compound.

All nine observed features in the experimental absorption spectrum have been marked from B$_1$ to B$_9$. All the calculated spectra corresponding to NBCO 5, 55, and 575 have nine features as shown in Fig. \ref{XANES_Ba}(b). On comparison among the calculated spectra, we observe that all the spectra show three features corresponding to B$_3$, B$_4$, and B$_5$ above the B$_2$ feature, in line with the experimental spectrum. But when we look closely, these features have different peak intensities in comparison to each other, depending upon the superstructure, as shown in the inset of Fig. \ref{XANES_Ba}(b). In the experimental spectrum, these three features can clearly be seen in the inset of Fig. \ref{XANES_Ba}(d). Based on the peak intensity ratio, the experimental spectrum resembles the NBCO 55 spectra more closely. 

Upon careful examination of features B$_5$ to B$_8$, as illustrated in Figs. \ref{XANES_Ba}(c) and (c'), it is evident that, despite the peak locations not being precisely aligned, all three superstructures exhibit four distinct features. The intensity ratio of features B$_5$ and B$_6$ in the experimental spectrum aligns more closely with the NBCO 575 spectrum. Given that the $\delta$ for the sample under investigation (0.65) lies between 0.55 and 0.75, the observations regarding the Ba L$_{\textnormal{III}}$ edge are coherent and consistent with the XRD and Nd L$_{\textnormal{III}}$ edge findings.

Under the dipole approximation, the allowed electronic transitions at the Ba $L_{\text{III}}$ edge involve excitations from the Ba $2p_{3/2}$ core level to unoccupied Ba $5d$- and $6s$-derived states. Fig. \ref{XANES_Ba}(d) displays the experimental Ba $L_{\text{III}}$ near-edge structure spectrum, while Fig. \ref{XANES_Ba}(e) shows the calculated unoccupied partial density of states (PDOS) for the Ba $5d$ states. The Fermi level ($\epsilon_F$) for the experimental spectrum was calibrated to $5242.22\text{ eV}$ by aligning the white line peak with the maximum of the $5d$ PDOS. Spectral features $\text{B}_2$–$\text{B}_5$ in the experimental absorption curve – align well with the unoccupied Ba $5d$ PDOS [Figs.~\ref{XANES_Ba}(d) and (e)]. Although the Ba $2p_{3/2} \rightarrow \text{Ba } 6s$ transition probability is inherently low, the broad profile of features $\text{B}_1$ indicates that $6s$ states actively participate in the absorption process, as depicted in Fig.~\ref{XANES_Ba}(f). Figs.~\ref{XANES_Ba}(f) and (g) show Ba $6p$ and O $2p$ PDOS, respectively. In the energy region corresponding to feature $\text{B}_1$, both spectra exhibit minor contributions and are hybridized with Ba $5d$ states above $\epsilon_F$ (inset of Fig.~\ref{XANES_Ba}(g)). Furthermore, the Ba $6p$ and O $2p$ PDOS display strong hybridised features around $\text{B}_2$ and $\text{B}_3$, and their contribution to the Ba $5d$ states remains considerable around the $\text{B}_4$ and $\text{B}_5$ features. Our results demonstrate that features $\text{B}_2$–$\text{B}_5$ receive significant contributions from $p$ symmetry states alongside Ba $5d$ and $6s$ states, highlighting that the atomic character of mainly Ba 5\textit{d} gets modified due to its hybridisation with p-symmtery states. The $\text{DFT}+U$ calculations successfully capture and explain the profile of the Ba $L_{\text{III}}$ edge spectral features.




\begin{figure}

\includegraphics [width = 0.48\textwidth]{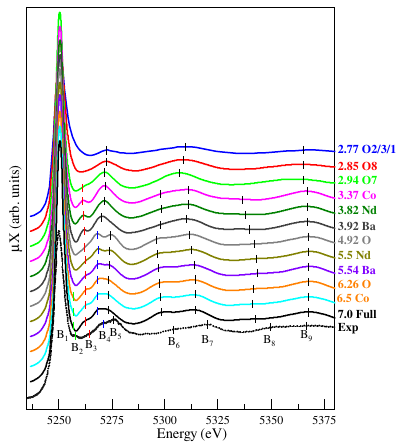}

\caption{Comparison of calculated Ba L$_\textnormal{III}$ absorption spectra for different cluster sizes with respect to the absorbing atom along with experimental spectra. The cluster size in \AA\ is given against each of the calculated spectra. The calculated spectra are shifted 4.5 eV towards lower energy, and vertical marks correspond to different features (B$_1$ - B$_9$) in the absorption spectra.}
\label{Var_Ba}

\end{figure}


To understand the origin of various features (B$_1$ - B$_8$) in Ba L$_{\textnormal{III}}$ absorption spectra in more detail, multiple scattering calculations have been performed using various cluster sizes around the Ba central atom in the NBCO 575 structure, as shown in Fig. \ref{Var_Ba}. Features B$_1$, B$_5$, B$_7$, and B$_9$ are found to develop with the inclusion of apical oxygens (O1, O2 and O3) in the Ba layer. While the intensity of these features increases with the inclusion of basal O8 and O7, feature B$_3$ starts to develop only after the incorporation of O7 in the cluster and develops further with the inclusion of other atoms. Features B$_6$ and B$_8$ start to develop only after the inclusion of Co atoms. This indicates that in the Ba L$_{\textnormal{III}}$ edge, Ba 5d states are modified due to their hybridisation with O 2p and Co 3d states giving rise to these features, in line with our DFT + \textit{U} results.

  The characteristic B$_4$ begins to appear only after the incorporation of second nearest neighbour O atoms, i.e., O5, O6 (apical), O7, O8 (basal) within the Nd layer and evolves with the addition of other atoms above 4.92 $\AA$. This feature is related to in-plane oxygens in the oxygen-deficient Nd layer. The features B$_7$ and B$_8$ also exhibit growth in intensity following the incorporation of the second nearest O atoms, located approximately 4.9 $\AA$ away.

\subsubsection{Co K-edge}

\begin{figure}

\includegraphics [width = 0.48\textwidth]{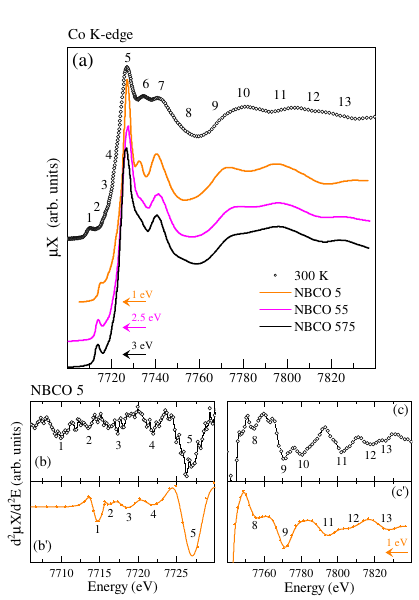}

\caption{ Comparison of (a) the normalised experimental Co K-edge absorption spectrum at 300 K with the calculated spectra for NBCO 5, NBCO 55, and NBCO 575, which have been shifted 1 eV, 2.5 eV, and 3 eV, respectively, to match the white line in the experimental spectra. Second derivative of (b) experimental pre-edge and (b') calculated NBCO 5 spectra; (c) experimental near-edge spectrum and (c') calculated NBCO 5 spectrum.}
\label{XANES1}

\end{figure}


In Fig. \ref{XANES1}(a), we present the normalised Co K-edge experimental x-ray absorption spectrum covering the pre-edge, edge and near-edge regions of NdBaCo$_2$O$_{5+\delta}$, where $\delta \sim$ 0.65, at room temperature (300 K). The features were labelled based on the second derivative minima of the experimental spectrum, Fig. \ref{XANES1}(b) and (c). Here, we observe 13 features.




Results of our calculations show that with an increase in the value of $\delta$, the intensity of the white line (feature 5)  is found to decrease, as shown in Fig. \ref{XANES1}(a). The ratio of the intensity of features 5 to 7, i.e., I(5)$:$I(7), for the experimental spectrum does not accurately correspond to any of the calculated spectra associated with varying non-stoichiometry. However, the calculated spectra corresponding to $\delta$ = 0.75 exhibit the lowest ratio that appears close to the experimental one. 



We now compare the second derivative of experimental spectrum with the calculated spectrum of NBCO5 in the pre-edge, Figs. \ref{XANES1}(b) and (b'), and XANES regions, Fig. \ref{XANES1}(c) and (c'). Our results show that the experimental spectrum shows 5 features (1-5) in the pre-edge region. The calculated spectrum corresponding to NBCO 5, where all the Co ions are in a square pyramidal environment, shows five features in the pre-edge region, as shown in Fig. \ref{XANES1}(b'). In the near-edge region, the experimental spectrum exhibits six distinct features (8-13). The calculated spectrum for NBCO 5 exhibits only five features in this region, while no feature is observable corresponding to feature 10, as shown in Fig. \ref{XANES1}(c'). These results show that the absorption spectrum corresponding to NBCO 5 can't explain the experimental spectrum satisfactorily.


\begin{figure}

\includegraphics [width = 0.48\textwidth]{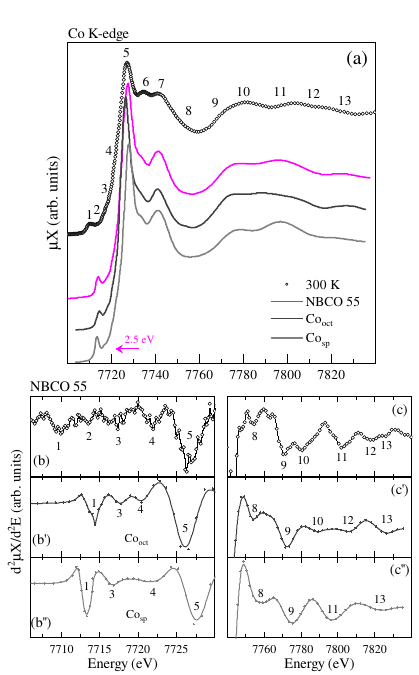}

\caption{Comparison of (a) normalised experimental Co K edge at 300 K for the sample under study along with calculated spectra for NBCO 55 and different Co environments, i.e., octahedral (Co$_{oct}$) and square pyramidal (Co$_{sp}$). Second derivative of (b) experimental pre-edge at 300 K along with calculated spectra for (b') Co$_{oct}$ and (b'') Co$_{sp}$. Second derivative of (c) experimental near-edge spectrum at 300 K along with calculated spectra for (c') Co$_{oct}$ and (c'') Co$_{sp}$. The calculated spectra have been shifted 2.5 eV to match the white line in the experimental spectra.}
\label{XANES2}

\end{figure}

NBCO 55 has two distinct Co environments due to oxygen vacancy ordering, with half of the Co ions in an octahedral environment and the rest in a square pyramidal environment. In Fig. \ref{XANES2}(a), we compare the second derivative of the experimental spectrum with the calculated spectrum of NBCO55 in the pre-edge and XANES regions for the Co ions in both environments, i.e., Co$_{oct}$ and Co$_{sp}$.


Our results show that the pre-edge region corresponding to both Co environments has only four features, as shown in Figs. \ref{XANES2}(b') and (b''). In the near-edge region, Co$_{oct}$ has five features, while Co$_{sp}$ has only four features in the calculated spectrum, as shown in Figs. \ref{XANES2}(c') and (c''), respectively. While we can accommodate all the features in the near-edge region (8-13) based on relative peak position. The pre-edge region (1-5) only has four features in addition to an intense white line in both Co environments. These results show that the calculated absorption spectrum corresponding to the NBCO 55 crystal structure also cannot satisfactorily explain the experimental spectra.

\begin{figure}

\includegraphics [width = 0.48\textwidth]{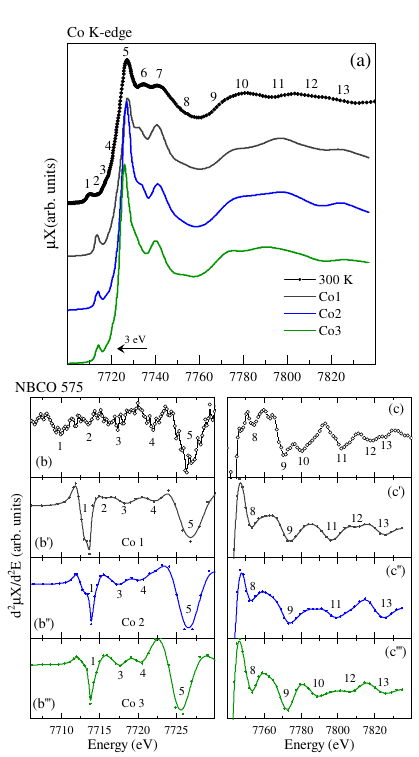}

\caption{Comparison of (a) normalised experimental Co K edge at 300 K with calculated and shifted absorption spectra for NBCO 575 with different Co central ions; double derivative of (c) experimental 300 K spectra along with calculated spectra taking different Co central ions. Second derivative of (b) experimental pre-edge spectrum at 300 K along with calculated spectra for different Co central ions in NBCO 575 (b') Co1, (b'') Co2 and (b''') Co3. Second derivative of (c) experimental near-edge spectrum at 300 K along with calculated spectra for (c') Co1, (c'') Co2 and (c''') Co3. The calculated spectra have been shifted 3 eV to match the white line in the experimental spectra.}

\label{XANES3}

\end{figure}

Fig. \ref{XANES3}(a) shows the comparison of the experimental spectrum with calculated spectra for the Co K edge in three distinct Co environments corresponding to the NdBaCo$_2$O$_{5.75}$ (NBCO 575) tetragonal (P4/mmm) crystal structure, i.e., Co1 square pyramidal (sp) and Co2 and Co3 octahedral (oct). As per our diffraction results, this structure is closest to the estimated non-stoichiometry in the compound. Initial observations reveal that the spectrum associated with Co1 (sp) closely corresponds to the experimental spectrum near the edge. The ratio of intensity for features 5, 6, and 7 is comparable to the experimental spectra. In the case of the spectra for Co2 and Co3 (oct), feature 5 is distinctly sharp. While no individual  Co environment can fully account for the entire experimental spectrum, averaging them also yields an unsatisfactory comparison, shown in Fig. \ref{XANES1}(a). It is noteworthy here that the stoichiometry of the examined sample is lower (0.65) than that of the model used (0.75) for Feff calculations; hence, it is inherently more probable that the spectra exhibit a greater square pyramidal character, which is indeed observed.



To examine the pre-edge and near-edge regions more closely, we have plotted the second derivative of the experimental spectrum, as shown in Fig. \ref{XANES3}(b) and (c). In the pre-edge region, Co1 spectrum show five features, whereas Co2 and Co3 spectra show four features.  In the near-edge region, we were also able to accommodate all the features (8 to 13) based on the relative peak positions in the second derivative of the calculated spectra for different Co ions, as shown in Fig. \ref{XANES3}(c'-c'''). Since all the pre-edge, edge and near-edge features can be calculated using the tetragonal crystal structure corresponding to $\delta$ = 0.75, this indicates that the local structure belonging to the tetragonal (P4/mmm) structure with a 222 superstructure is nearest to the experimental spectrum.

\begin{figure}

\includegraphics [width = 0.4\textwidth]{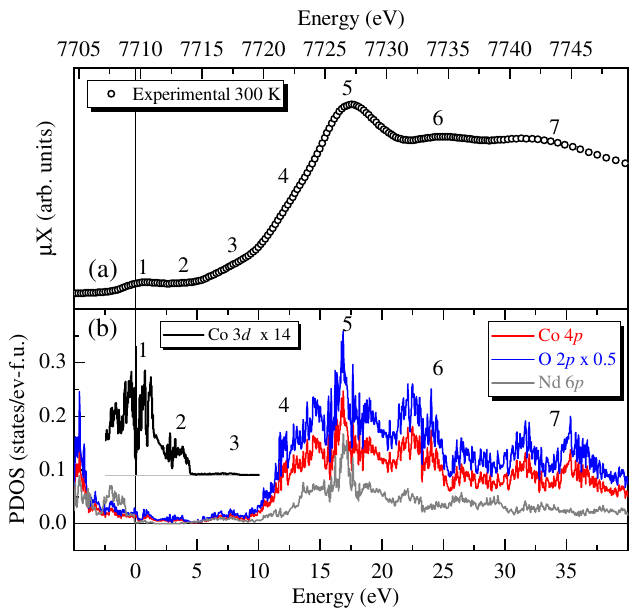}

\caption{ Experimental X-ray absorption spectrum for Co K-edge collected at 300 K in the pre-edge and edge regions. Partial density of states (PDOS) of unoccupied states of for Co 3\textit{d}, Co 4\textit{p}, O 2\textit{p} and Nd 6\textit{p} from DFT+U calculation for NBCO 575; The inset shows Co 3\textit{d} PDOS spectrum multiplied by 14 near features 1-3, and O 2\textit{p} PDOS spectrum is divided by 2 in the figure.}

\label{XANES_2}

\end{figure}


Fig. \ref{XANES_2}(b) displays the partial density of states (PDOS) for Co 4\textit{p}, Co 3\textit{d}, O 2\textit{p}, and Nd 6\textit{p} states in the unoccupied energy region above the Fermi level ($\epsilon_F$). In the dipole approximation, the K-edge represents the transition of an electron from the 1\textit{s} state to unoccupied \textit{p} states (Co 4\textit{p}). Features 4 and 5 at the edge are dominated by Co 4\textit{p} states. Apart from Co 4\textit{p} states, there is also a strong contribution of O 2\textit{p} and a weak contribution of Nd 6\textit{p} states. This suggests that strong and weak hybridisation of O 2\textit{p} and Nd 6\textit{p} states  with Co 4\textit{p} states, respectively, leads to the reduction in the atomic character of Co 4\textit{p} states.



According to the quadrupole approximation, transition from the 1\textit{s} state to unoccupied \textit{d} (Co 3\textit{d}) states in transition metals results in pre-edge structure in absorption spectra. In features 1, 2, and 3, we see a high density of states (DOS) for Co 3\textit{d}, which are strongly hybridised with O 2\textit{p} states. This suggests that features 1, 2, and 3 are dominated by Co 3\textit{d} states, with substantial \textit{p}-\textit{d} hybridisation in these oxygen-deficient double perovskites. Such \textit{p}-\textit{d} admixture essentially leads to an increment in the probability of transition in the pre-edge region. Fig. \ref{XANES_2}(a) shows pre-edge and edge regions of the absorption spectrum with respect to energy. The features 6 and 7 arise from the modification in the atomic character of Co 4\textit{p} states as a result of their strong and weak hybridisation with O 2\textit{p} and Nd 6\textit{p} states, respectively. These results show that the DFT + \textit{U} calculations give a better representation of the experimental spectrum, as shown in Figs. \ref{XANES_2}(a) and (b).




\begin{figure*}

\includegraphics [width = 1.0\textwidth]{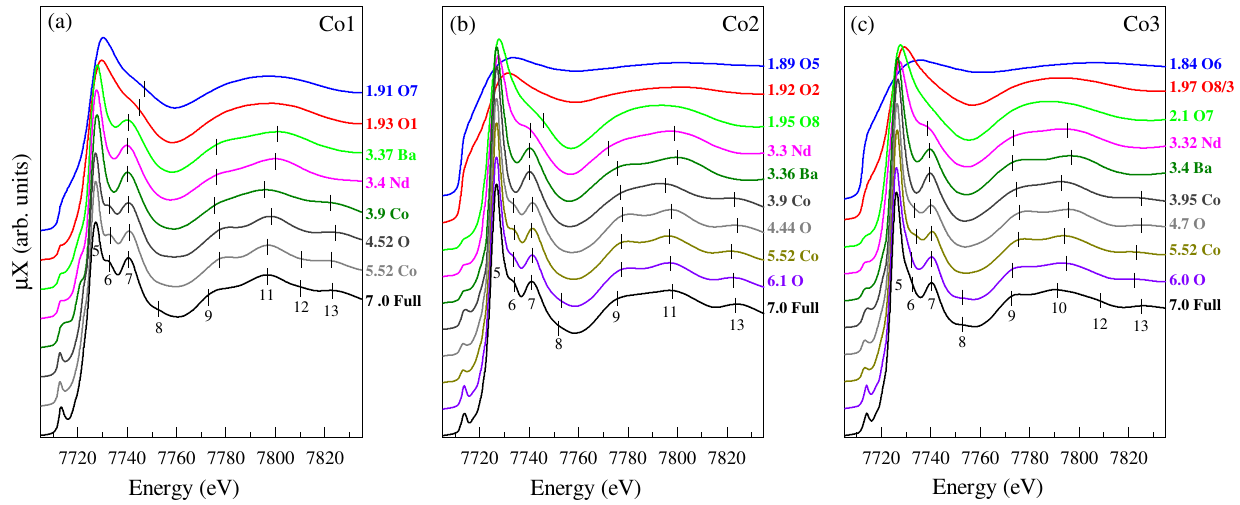}

\caption{Calculated spectra for different cluster sizes with absorbing atoms as (a) Co1; (b) Co2 and (c) Co3. The number in front indicates the distance of scatterer in \AA\, from the absorbing atom and scatterer/s situated at that distance. The calculated spectra are shifted 3 eV towards lower energy and vertical lines are corresponding to different features (5-13) as observed after doing the double derivative above the absorption edge, shown in Fig. \ref{XANES1}(a). }
\label{Var_Co}

\end{figure*}

To further understand the origin of different features present in the spectrum (5-13), multiple scattering calculations have been performed for various cluster sizes taking  Co1, Co2 and Co3 as the central atom, as shown in Figs. \ref{Var_Co} (a-c). These calculations were done using the tetragonal (P4/mmm) 222 structure with a  $\delta$ = 0.75. The cluster size was varied considering different scattering atom/s upto 7 \AA\, in these calculations. We observe that the white line (feature 5) in the case of Co1, Co2 and Co3 prominently starts to increase only after the inclusion of the basal O7 and O8 atoms in the calculations; that can be clearly seen for the case of Co2 and Co3, as shown in Fig. \ref{Var_Co} (a-c). 


It is interesting to note here that the intense white line observed in the calculated spectrum is mainly due to the inclusion of basal O7 and O8 atoms in the cluster. However, the intensity of the white line of the experimental spectrum is quite low. Such low intensity could be due to the vacancy-generated discrepancy in the oxygen ions from some of the O7 and O8 sites in the basal plane, in addition to the vacancy in the oxygen at the apical site, as the prepared sample is more oxygen deficient ($\delta$ = 0.65) as compared to the calculated spectrum ($\delta$ = 0.75). To the best of our knowledge, among the possible superstructures available in literature for non-stoichiometry $\delta <$ 0.75, $\delta$ = 0.50 is well studied, where the extra oxygen goes away from the apical positions in the Nd layer, giving rise to 122 superstructure investigated using neutron and synchrotron diffraction studies\cite{burley2003, fauth2002spin, maignan1999}. Other than that, the common consensus for RBaCo$_2$O$_{5.5}$ (R other than Nd) systems also shows 122 superstructure\cite{taskin2003, moritomo2000metal, akahoshi2001oxygen, miao2017hole, galakhov2022insight}. As previously discussed, in our multiple scattering calculations, it can also be seen that the $\delta$ = 0.50 system has a higher intensity of the white line in comparison to $\delta$ = 0.75, which is not in line with the experimental observation, as shown in Fig. \ref{XANES1}(a).  However, to ascertain the amount and type of oxygen vacancies, high-resolution neutron diffraction experiments will be helpful, where the Wyckoff positions corresponding to basal and apical oxygens can be individually refined to get the $\%$ vacancies on different sites.



 We observe that the feature 6 starts to increase after the inclusion of the second nearest neighbour O atoms in the system. While for the spectrum with Co1 and Co2 central atoms, this feature can be clearly seen, for Co3 as the central atom, this feature is quite broad. The intensity of feature 7 shows a prominent increase after the inclusion of Ba and Nd atoms, which indicates that this feature is related to the transition to modified Co 4\textit{p} states due to hybridisation with Nd and Ba 6 \textit{p} symmetry states.

Features 8, 9 and 13 are the common features obtained in the calculations for all three Co environments, and feature 12 is observed in the case of Co1 and Co3, as previously seen in Fig. \ref{XANES3}(c). Features 8 and 12 are clearly apparent only after the calculation with the full cluster (7$\AA$). Feature 9 first appears after adding the Ba atoms, and feature 13 starts to appear after adding Co ions in the calculation. There are three remaining features; one appears in the case of Co1 and Co2 (feature 11) and one in the case of Co3 (feature 10).  In Fig. \ref{Var_Co}(c), we can observe that the broad feature around 7800 eV that first appears after adding Nd atoms in the calculation is present at $\sim$ 3.32 $\AA$ and then continuously shifts towards lower energy, giving rise to feature 10, which is unique to Co3. Similarly, in Figs. \ref{Var_Co}(a) and (b), we observe one feature around 7800 eV (feature 11) growing after adding Nd atoms in the calculation present $\sim$ 3.32 $\AA$ for Co1 and Co2, but after adding other atoms, the shift in this feature is minimal, giving rise to feature 11. Overall, the local structure study using Co K-edge and multiple scattering calculations shows that to reproduce the whole experimental spectra, all three kinds of Co ions are required, indicating the stabilisation of the local structure of the 222 superstructure in the compound and also indicating that it is possible that some of the oxygen might be absent from the basal position in the case of the sample under study, i.e., NdBaCo$_2$O$_{5.65}$. 

The additional basal oxygen vacancies are expected to directly affect the in-plane Co–O–Co magnetic exchange network, producing much stronger magnetic disorder and potentially enhancing Griffiths-like behaviour and suppression of long-range magnetic order\cite{Pant_2025}. Additionally, such vacancies can lead to localisation centres, thereby leading to a variable range hopping transport mechanism, as observed in this material\cite{Pant_2025}. From the application point of view, until an optimum value, the basal vacancies can connect vacancy channels already created by apical vacancies, resulting in easier oxygen diffusion and potentially better ionic conductivity.


\section{Conclusion}

A polycrystalline sample of $\text{NdBaCo}_2\text{O}_{5+\delta}$ ($\delta \sim$ 0.65) has been synthesised via the auto-combustion method. This compound has been studied using X-ray absorption near edge structure (XANES) spectroscopy at 300 K, supplemented by multiple scattering calculations and first-principles calculations based on density functional theory ($\text{DFT}+U$). The present results show a direct link  between oxygen-vacancy distribution, magnetic inhomogeneity, and charge transport in layered cobaltates. The analysis for Co K and Nd L$_{\text{III}}$ edges reveals that the local structure of the compound under study matches the spectral features of the NBCO 575 structure. The shape of the different absorption edges aligns well with the calculated density of states (DOS), yielding vital information about the hybridisation between the absorbing atoms and their surrounding ions. The +2 oxidation state for Ba ions is confirmed by the Ba L$_{\text{III}}$ spectra, while the shape and number of spectral features in the Ba L$_{\text{III}}$ near-edge region further suggest that the $\delta$ value for the compound lies between 0.5 and 0.75, which corroborates well with the actual composition of our sample. A more thorough investigation of the Co K edge through calculations of various cluster sizes suggests that the additional oxygen vacancies may arise at the basal sites; this contradicts the common assumption that oxygen is typically missing from the apical sites in the Nd layer of these compounds. We also provide a microscopic origin of the observed Griffith phase and variable range conduction mechanism exhibited by the compound. The observation of oxygen vacancies at both apical and basal sites suggests the possibility of its utility for a sensitive oxygen sensor when the vacancies are at the optimum level.

\section{Acknowledgment}

This research used beamline QAS of the National Synchrotron
Light Source II, a U.S. Department of Energy (DOE) Office of
Science User Facility operated for the DOE Office of Science
by Brookhaven National Laboratory.

\bibliography{Manuscript}






\end{document}